\documentclass[runningheads]{llncs}
\usepackage{graphicx}
\usepackage{hyperref}

\usepackage{tikz}
\usepackage{amsmath}
\usepackage{filecontents}
\usepackage{graphics}
\usepackage{amsmath}
\usepackage{multirow}
\usepackage{makecell}
\usepackage{algorithm}
\usepackage{algpseudocode}
\usepackage{pifont}
\usepackage{booktabs}
\usepackage{amssymb}
\usepackage{adjustbox}
\usepackage{multirow}
\usepackage{caption}
\usepackage{subcaption}

\newcommand{\sysname}{AttacKG }

\newcommand{\highlightb}[1]{{\color{black}{#1}}}

\definecolor{patriarch}{rgb}{0.5, 0.0, 0.5}
\definecolor{fncolor}{rgb}{0.0, 0.42, 0.24}
\definecolor{fpcolor}{rgb}{0.77, 0.12, 0.23}

\newcommand{\cmark}{\ding{51}}%
\newcommand{\xmark}{\ding{55}}%

\begin{document}
\title{AttacKG: Constructing Technique Knowledge Graph from Cyber Threat Intelligence Reports}
\titlerunning{AttacKG: Constructing Technique Knowledge Graph from CTI Reports}

\author{
Zhenyuan Li\inst{1} \and
Jun Zeng\inst{2} \and
Yan Chen\inst{3}\and
Zhenkai Liang\inst{2} 
}
\authorrunning{Z. Li et al.}
% \authorrunning{Z. Li, J. Zeng, Y. Chen, and Z. Liang}
\institute{
Zhejiang University, Hangzhou, China\\ 
% \email{lizhenyuan@zju.edu.cn}
\and
National University of Singapore, Singapore\\
% \email{\{junzeng,liangzk\}@comp.nus.edu.sg}
\and
Northwestern University, Evanston, USA\\
% \email{ychen@northwestern.edu}
}

\maketitle              

\vspace{-0.1in}
\begin{abstract}
Cyber attacks are becoming more sophisticated and diverse, making detection increasingly challenging. To combat these attacks, security practitioners actively summarize and exchange their knowledge about attacks across organizations in the form of cyber threat intelligence (CTI) reports. However, as CTI reports written in natural language texts are not structured for automatic analysis, the report usage requires tedious manual efforts of cyber threat intelligence recovery. Additionally, individual reports typically cover only a limited aspect of attack patterns (techniques) and thus are insufficient to provide a comprehensive view of attacks with multiple variants.

To take advantage of threat intelligence delivered by CTI reports, we propose AttacKG to automatically extract structured attack behavior graphs from CTI reports and identify the adopted attack techniques. We then aggregate cyber threat intelligence across reports to collect different aspects of techniques and enhance attack behavior graphs into technique knowledge graphs (TKGs). 

In our evaluation against 1,515 real-world CTI reports from diverse intelligence sources, AttacKG effectively identifies 28,262 attack techniques with 8,393 unique Indicators of Compromises (IoCs). To further verify the accuracy of AttacKG in extracting threat intelligence, we run AttacKG on 16 manually labeled CTI reports. Empirical results show that AttacKG accurately identifies attack-relevant entities, dependencies, and techniques with F1-scores of \highlightb{0.887, 0.896, and 0.789, which outperforms the state-of-the-art approaches Extractor~\cite{Satvat2021} and TTPDrill~\cite{Husari2017}.
Moreover, the unique technique-level intelligence will directly benefit downstream security tasks that rely on technique specifications, e.g., APT detection and cyber attack reconstruction.}

% We evaluate AttacKG on 1,515 real-world CTI reports. 

% The large-scale experiments demonstrate that AttacKG can effectively extract technique-level threat intelligence from massive CTI reports. 

% Specifically, we identify 28,262 techniques available in 1,515 CTI reports with 8,393 unique IoCs. 

% Furthermore, eight reports are selected and manually labeled to evaluate \sysname's accuracy of attack graph extraction and technique identification. 

% Empirical results show that \sysname accurately extracts attack-relevant entities, dependencies, and techniques with F1-scores of 0.895, 0.911, and 0.819, which significantly outperforms state-of-the-art approaches like EXTRACTOR~\cite{Satvat2021} and TTPDrill~\cite{Husari2017}.

\end{abstract}

\section{Introduction}
\label{sec:introduction}

Advanced cyber security attacks are growing rapidly. 
The trend of cyber attacks is to adopt increasingly sophisticated tactics and diverse techniques~\cite{SolarWindshack}, such as multi-stage Advanced Persistent Threats (APTs), making detection more challenging than ever before. 
To combat against these attacks, security analyzer actively exchange threat intelligence to enhance their detection capabilities.

% On the one hand, (traditional) IoC-based threat intelligence only cover limited attack tactic and easy-to-bypass. build on strong assumption.

\highlightb{Among them, structured threat intelligence defined by open standards such as OpenIoC~\cite{OpenIOC}, STIX~\cite{STIX}, and CybOX~\cite{CybOX} are widely shared on open-source platform, such as AlienVault OTX\cite{alienvault} and IBM X-Force\cite{xforce} and utilized in security operation centers. }
Such intelligence standards defines cyber attacks as IoCs, which are artifacts in forensic intrusions, such as MD5 hashes of malware samples and IP/domains of command-and-control (C\&C) servers.
However, recent studies have shown that detection with such disconnected IoCs is easy to bypass~\cite{Li2019,Milajerdi2019}. 
For example, attackers can frequently change domains used in attack campaigns to evade detection. 
In comparison, by taking IoC interactions into account, graph-based detection typically demonstrates better robustness~\cite{MomeniMilajerdi2019} by capturing attack techniques aligned to adversarial goals. 
Specifically, attack techniques~\cite{hossain2020combating,mitre} are basic units that describe ``how'' attack actions are performed and are widely reused in different attack campaigns. 

Such technique implementations details can be found in unstructured CTI reports written by security practitioners based on their observations of attack scenarios in the wild. 
In particular, a well-written report will precisely describe attack behaviors through enumerating attack-relevant entities (e.g., \textit{CVE-2017-11882}) and their dependencies (e.g., \texttt{stager} connecting to \texttt{C\&C sever}). 
However, recovering attack behaviors from textual CTI requires non-trivial manual efforts.
Intuitively, a system capable of automatically extracting attack techniques knowledge from CTI reports can significantly benefit cyber defenses by reducing human efforts and speeding up response time. 
We identify two main challenges in expert knowledge extraction from CTI reports: 
(1) As CTI reports are written in an informal format, in natural languages, extracting structured attack behaviors needs to analyze semantics in unstructured CTI texts; 
(2) Attack knowledge is dispersed across multiple reports. Individual reports commonly only focus on limited/incomplete attack cases, making it difficult to obtain a comprehensive view of attacks. 
And existing works on CTI report parsing~\cite{Satvat2021,Husari2017,ghazi2018supervised,husari2018using,ramnani2017semi} only focus on parsing attack cases within single reports.

In this paper, we propose \sysname, a novel approach to aggregate threat intelligence across multiple CTI reports for attack techniques and construct a knowledge-enhanced attack graph summarizing the technique-level attack workflow in CTI reports. Based on enhanced knowledge, we introduce a new concept which we call ``Technique Knowledge Graph'' (TKG), that summarizes causal techniques from attack graphs to describe the complete attack chain in CTI reports. 
% \highlightb{A relevant concept is ``Tactic Provenance Graph'' (TPG) proposed by Hassan \textit{et, al.} \cite{Hassan2020}.}
More specifically, we first adopt a pipeline to parse a CTI report and extract attack-relevant entities and entity dependencies as an attack graph. Then, we initialize technique templates using attack graphs built upon technique procedure examples crawled from the MITRE ATT\&CK knowledge base~\cite{mitre}. Next, we utilize a revised graph alignment algorithm to match technique templates in attack graphs. Towards this end, we can accurately align and refine the entities in both CTI reports and technique templates. While technique templates aggregate specific and probably new intelligence from real-world attack scenarios described in CTI reports, attack graphs can utilize such knowledge in templates to construct technique knowledge graphs (TKGs).

We implement \sysname and evaluate it against 7,373 examples of 179 techniques crawled from MITRE ATT\&CK and 1,515 CTI reports collected from multiple intelligence sources~\cite{ciscotalos,darpatc}. % ,microsoft
% such as Cisco Talos Intelligence Group~\cite{ciscotalos}, Microsoft Security Intelligence Center~\cite{microsoft}, and DARPA Transparent Computing (TC) program~\cite{darpatc}. 
Our experimental result demonstrates that \sysname substantially outperforms existing CTI parsing solutions such as EXTRACTOR~\cite{Satvat2021} and TTPDrill~\cite{Husari2017}: 
(1) With our CTI report parsing pipeline, \sysname accurately constructs attack graphs from reports with F1-scores of \highlightb{0.887} and \highlightb{0.896} for entities and dependencies extraction, respectively;
(2) Based on extracted attack graphs, \sysname accurately identifies adversarial techniques  with an F1-score of \highlightb{0.789};
(3) \sysname can successfully collect 28,262 techniques, and 8,393 unique IoCs from 1,515 CTI reports.

To the best of our knowledge, this is the first work to aggregate attack knowledge from multiple CTI reports at the technique level. 
In particular, our work makes the following contributions:

\vspace{-\topsep}
\begin{list}{\labelitemi}{\leftmargin=1.5em}
 \setlength{\topmargin}{0pt}
 \setlength{\itemsep}{0em}
 \setlength{\parskip}{0pt}
 \setlength{\parsep}{0pt}
    \item We present a new pipeline for CTI report parsing with better \highlightb{efficiency and} performance in handling co-reference and constructing attack graphs.
    
    \item We propose the design of the technique template to describe and collect general technique knowledge, and a revised graph alignment algorithm to identify attack techniques with templates. By aligning templates with the technique implementation in the attack graph, we exchange the knowledge from both to refine each other and form technique knowledge graphs (TKGs).
    
    \item We implement AttacKG (open sourced\footnote{To facilitate follow-up research, we release the source code of AttacKG at \url{https://github.com/li-zhenyuan/Knowledge-enhanced-Attack-Graph}.}) and evaluate it with 1,515 real-world CTI reports. The result demonstrates that we can extract attack graphs from reports accurately and aggregate technique-level threat intelligence from multiple unstructured CTI reports effectively. 
    \highlightb{And we discuss the benifit of TKGs with two case studies.}
\vspace{-\topsep}
\end{list}

\section{Background and Related Work}
\label{sec:background}

In this section, 
\highlightb{
We first introduce the outbreaking attack mutations.
% We first present how the knowledge in Cyber Threat Intelligence (CTI) reports benefits threat detection.
Then, we introduce state-of-the-art threat intelligence extraction methods.
Finally, we present a real-world CTI report as a motivating example for intuitive illustration.
}

\highlightb{ 
\vspace{-0.25in}
\subsection{Cyber Attacks and Reports}
\label{subsec:mutation}

Attackers actively create attack variants to bypass detection.
To systematize and summarize the behavior in the attack variants, MITRE proposed the ATT\&CK Tactics-Techniques-Procedures (TTP) matrix based on real-world observations of cyber attacks.
In the hierarchical TTPs matrix, tactics describe ``why'' given adversarial action is performed, which are typically fixed for an attack, while the selection and implementation of techniques that describe ``how'' to perform adversarial action is more flexible. 

% \vspace{-0.15in}
\begin{figure*}[htbp]
    \centering
    \includegraphics[width=1\textwidth]{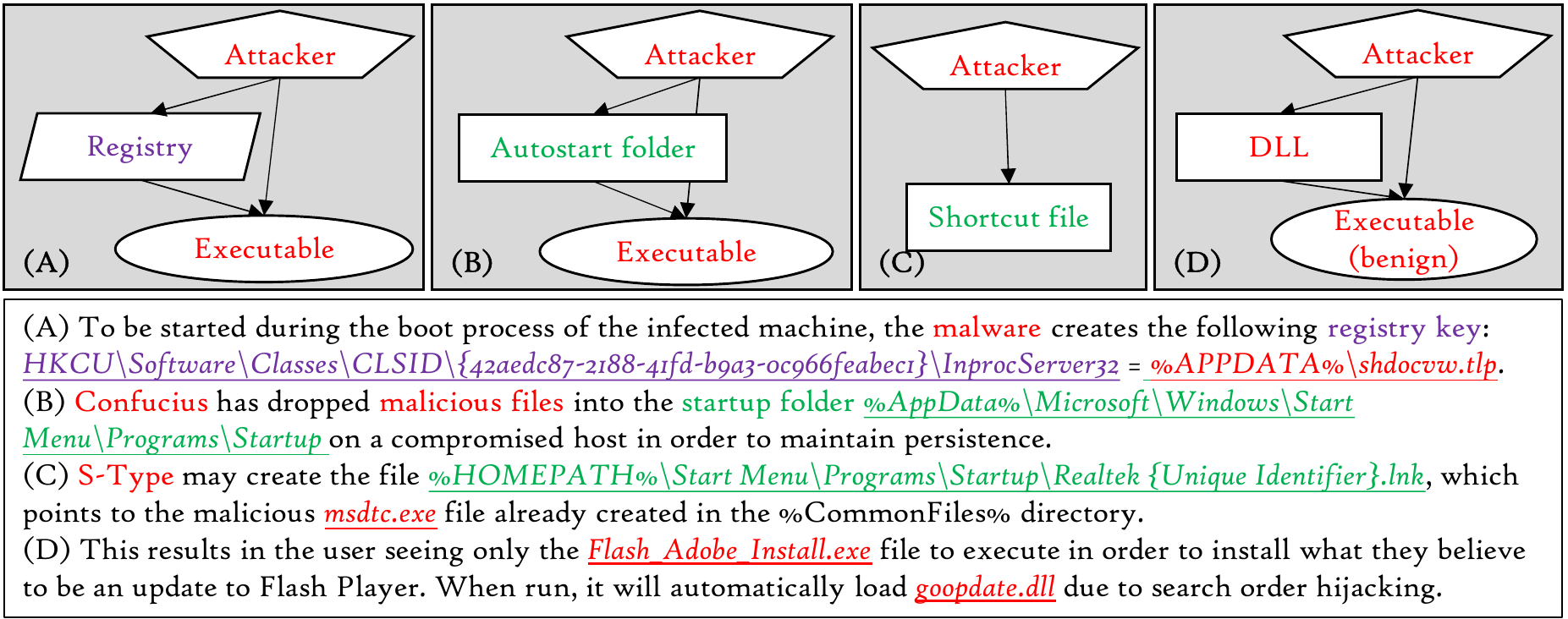}
    \vspace{-0.2in}
    \caption{\highlightb{\small{Technique template generated by \sysname and corresponding real-world description for ``\textit{T1547-Boot or Logon Autostart Execution}'' with four variants corresponding to fourteen MITRE sub-techniques categorized as (A) \textit{Registry Run Keys}, (B) \textit{Auto-start folder}, (C) \textit{Shortcut Modification}, and (D) \textit{DLL Side-loading}.}}}
    \label{fig:template}
\vspace{-0.2in}
\end{figure*}

As Figure~\ref{fig:template} shows, the commonly used technique ``\textit{T1547-Boot or Logon Autostart Execution}'' for tactic ``\textit{Persistent}'' can be implemented in at least four different ways: (A) ``\textit{Registry Run Keys}'', (B) ``\textit{Auto-start folder}'', (C) ``\textit{Shortcut Modification}'', and (D) ``\textit{DLL Side-loading}''. 
The number of variants grows exponentially if the implementation details, such as the selection of registry key, are taken into consideration.
Our observation is that the ways of technique implementation are relatively limited, while the selection of implementation details is much more varied.
Therefore, it is reasonable to believe that a system that can identify various techniques and collect implementation details would significantly benefit downstream security tasks by providing practicable intelligence.

The manually-crafted TTP matrix cannot cover the various implementations.
Such detailed technique implementation knowledge only comes from practice.
Thus, security practitioners actively gather and share knowledge about attacks as threat intelligence.
Such cyber threat intelligence (CTI) is managed and exchanged typically in the form of either structured and machine-digestible Indicators of Compromise (IoCs) or unstructured and natural language reports.
}

\vspace{-0.2in}
\subsection{Threat Intelligence Extraction}
% \vspace{-0.03in}
Cyber threat intelligence (CTI) plays a vital role in security warfare to keep up with the rapidly evolving landscape of cyber attacks~\cite{mu2018understanding,gao2021system}.
To facilitate CTI knowledge exchange and management, the security community has standardized open formats (e.g., OpenIoC~\cite{OpenIOC}, STIX~\cite{STIX}, and CybOX~\cite{CybOX}) to describe Indicators of Compromises (IoCs).
Though structured and machine-readable, such intelligence lacks semantic information about how entities/IoCs interact to form the kill chain. 
While in this paper, we try to fulfill this semantic gap with technique-level knowledge aggregated from various CTI reports.

\begin{table}[htbp]
\scriptsize
    \centering
    \vspace{-0.11in}
    \caption{\highlightb{Comparison of threat intelligence extraction methods}}
	\begin{tabular}{cccccc}
        \toprule
    	 & Automatic & Graph-structure & Technique-aware  & Cross-reports \\% & Generalization\\
    	 \midrule
        Poirot~\cite{Milajerdi2019} & \xmark & \cmark & \xmark & \xmark \\% & Regex \\ 
    	iACE~\cite{Liao} & \cmark & \xmark & \xmark & \xmark \\% & \xmark\\
    	Extractor~\cite{Satvat2021} \& ThreatRaptor~\cite{gao}  & \cmark & \cmark & \xmark & \xmark\\% & Over-generalized\\
    % 	ChainSmith~\cite{Zhu2018} & \cmark & \xmark & \cmark & \xmark\\% & \xmark\\  
    	TTPDrill~\cite{Husari2017} \& rcATT\cite{legoy2020automated}, etc. & \cmark & \xmark & \cmark & \xmark\\% & \xmark\\ 
    % 	rcATT\cite{legoy2020automated} & \cmark & \xmark & \cmark & \xmark\\% & \xmark\\ 
    	AttacKG & \cmark & \cmark & \cmark & \cmark\\% & Enumeration\\
    	\bottomrule
    \end{tabular}
    \label{tab:related_work}
    \vspace{-0.2in}
\end{table}

% \begin{table}[htbp]
% \scriptsize
%     \centering
%     \vspace{-0.11in}
%     \caption{{Comparison of threat intelligence extraction methods}}
% 	\begin{tabular}{cccccc}
%         \toprule
%     	 & Extractor & Poirot & TTPDrill & AttacKG \\
%     	 \midrule
%     	 Automatic & \cmark & \cmark & \xmark & \xmark \\
%          Graph-structure & \xmark & \cmark & \xmark & \xmark  \\  
%     	 Technique-aware & \cmark & \xmark & \cmark & \xmark \\ 
%     	 Cross-reports & \cmark & \cmark & \cmark & \cmark \\
%     	 Generalization & Over-generalized & Regex & \xmark & Enumeration\\
%     	\bottomrule
%     \end{tabular}
%     \label{tab:related_work}
%     \vspace{-0.2in}
% \end{table}

\highlightb{
Poirot~\cite{Milajerdi2019} utilizes manually extracted and generalized (attack) query graphs for intrusion detection in provenance graphs constructed from system logs, which validates the efficacy of threat intelligence for detection.
However, manually extracting attack-relevant information from uninstructed texts is labor-intensive and error-prone, hindering CTI's practice applications.
Therefore, several approaches have been proposed to analyze CTI reports automatically.
As Table~\ref{tab:related_work} shown, these works can be roughly divided into several categories.
Specifically, iACE~\cite{Liao} presents a graph mining technique to collect IoCs available in tens of thousands of security articles.
Extractor~\cite{Satvat2021} and ThreatRaptor~\cite{gao} customize NLP techniques to model attack behaviors in texts as attack graphs.
TTPDrill~\cite{Husari2017}, rcATT~\cite{legoy2020automated} and ChainSmith~\cite{Zhu2018} derive threat actions from reports and map them to attack patterns (e.g., tactic and techniques in MITRE ATT\&CK~\cite{mitre}) with pre-defined ontology or machine learning models.
Similar to prior studies, the large body of \sysname is to automate knowledge extraction from CTI.
Nevertheless, \sysname distinguishes itself from these works in the sense that it identifies TTPs and constructs technique knowledge graphs (TKGs) to summarize technique-level knowledge across CTI reports.
}

% While CTI reports provide detailed contexts about an attack (e.g., the sequence of adversarial actions), manually extracting attack-relevant information from uninstructed texts is labor-intensive and error-prone, hindering CTI's practice applications.
% Based on the observation that IoCs often share similar context terms, iACE~\cite{Liao} presents a graph mining technique to collect IoCs available in tens of thousands of security articles.
% ChainSmith~\cite{Zhu2018} further makes use of neural networks to classify IoCs into different attack campaign stages (e.g., baiting and C\&C).
% TTPDrill~\cite{Husari2017} derives threat actions from reports and maps them to attack patterns (e.g., techniques in MITRE ATT\&CK~\cite{mitre}) pre-defined as ontology.
% EXTRACTOR~\cite{Satvat2021} and ThreatRaptor~\cite{gao} customize NLP techniques to model attacks behaviors in texts as provenance graphs.

\vspace{-0.1in}
\subsection{Motivating Example}
\label{subsec:motivating}

\begin{figure*}[ht]
    \centering
    \vspace{-0.2in}
    \includegraphics[width=0.95\textwidth]{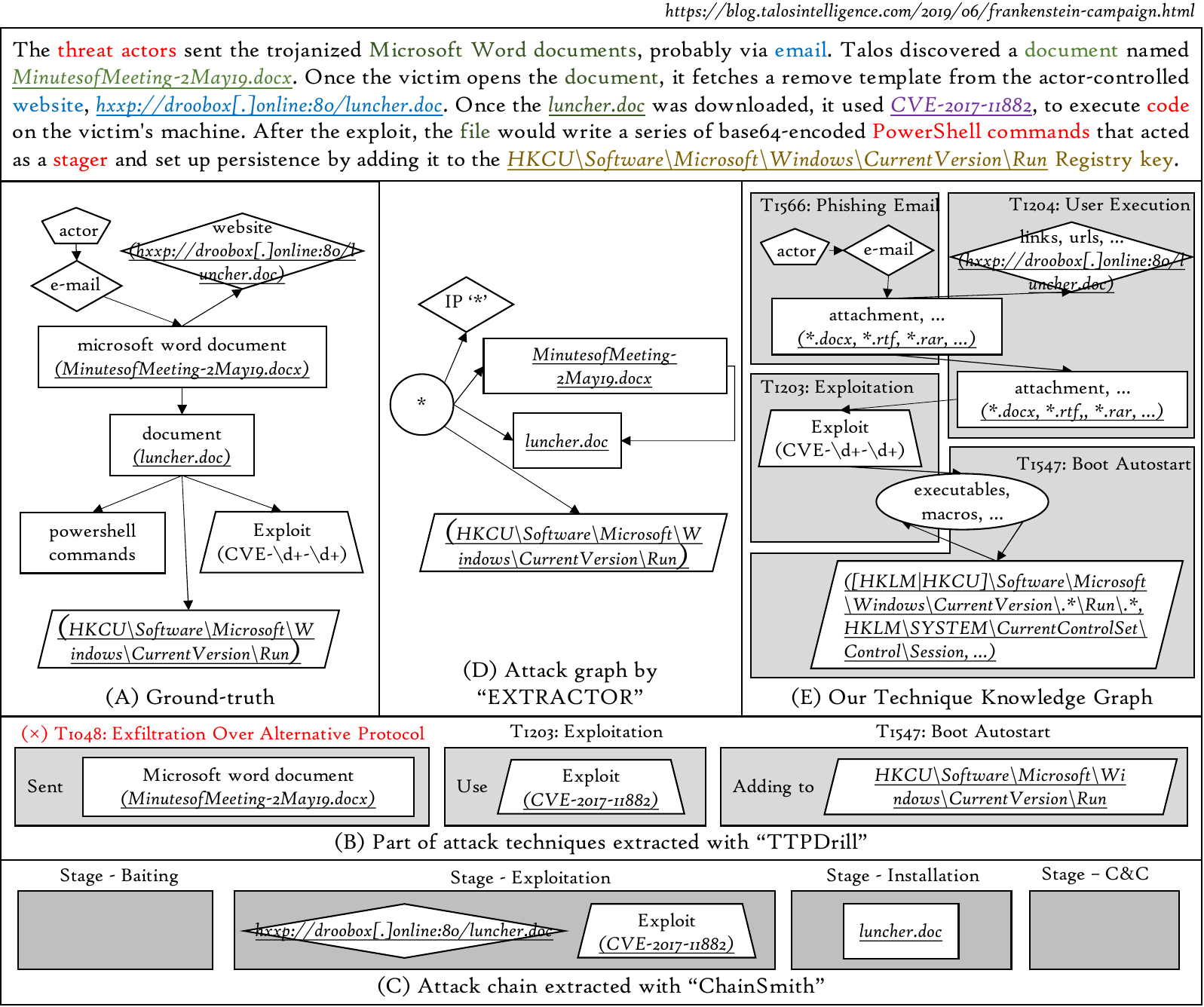}
    \vspace{-0.07in}
    \caption{A motivating example.}
    \label{fig:example}
    \vspace{-0.22in}
\end{figure*}

Figure~\ref{fig:example} presents a real-world APT attack campaign called Frankenstein~\cite{frankenstein}. 
The campaign name comes from the ability of threat actors to piece together different independent techniques.
As shown, this campaign consists of four attack techniques, namely, \textit{T1566-Phishing E-mail}, \textit{T1204-User Execution}, \textit{T1203- Exploitation} and \textit{T1547-Boot Autostart}.
Each technique involves multiple entities and dependencies to accomplish one or more tactical attack objectives.
It presents a typical multi-stage attack campaign that consists of multiple atomic techniques.
To evade detection, such attacks can be morphed easily by replacing any technique with an alternative one. 
Therefore, summarized knowledge of attack techniques, which is robust and semantically rich, is beneficial to detection and investigation of cyber attacks~\cite{Li2021,Hassan2020,Milajerdi2019,Husari2017}.

% \vspace{-0.2in}

% Research progress has been made to automatically extract knowledge about attacks from CTI reports.
Subfigures (B) to (D) show the attack knowledge retrieved from the report sample by TTPDrill~\cite{Husari2017}, ChainSmith~\cite{Zhu2018}, and EXTRACTOR~\cite{Satvat2021}, respectively, while Subfigure (A) represents the manually generated ground-truth.
Subfigure (B) shows attack techniques identified by TTPDrill with manual-defined threat ontology. As shown, TTPDrill can only extract separate techniques from CTI reports without the whole picture. Besides, the ontology provided by TTPDrill contains only action-object pairs for technique identification, which is too vague and may lead to numerous false positives.  As the example shows, sending a document is recognized as exfiltration in TTPDrill. However, the ``trojanized'' document is, in effect, sent by an attacker for exploitation.
% Detailed comparisons are demonstrated in Section~\ref{sec:evaluation}.
As shown in Subfigure (C), ChainSmith provides a semantic layer on top of IoCs that captures the different roles of IoCs in a malicious campaign. However, they only give a coarse-grained four-stage classification with limited information.
As Subfigure (D) shows, the attack graph generated by EXTRACTOR merges all non-IoC entities of the same type and thus loses the structural information of attack behaviors, making it impossible to identify the technique accurately.

Subfigure (E) illustrates the ideal result we would like to extract in this paper. As long as we can locate attack techniques in attack graphs extracted from CTI reports, we are able to aggregate technique-level knowledge and enrich the attack graphs with more comprehensive knowledge about the corresponding techniques. For example, we can find more possible vulnerabilities that can be used in \textit{T1203-Exploitation for Execution} as a replacement for \textit{CVE-2017-11882} appeared in this report. Moreover, the distinct threat intelligence can be collected and aggregated at the technique level across multiple CTI reports.

\vspace{-0.07in}
\section{Approach}
\label{sec:system_overview}

\subsection{Overview of \sysname}

\begin{figure*}[ht]
    \centering
    \vspace{-0.17in}
    \includegraphics[width=0.96\textwidth]{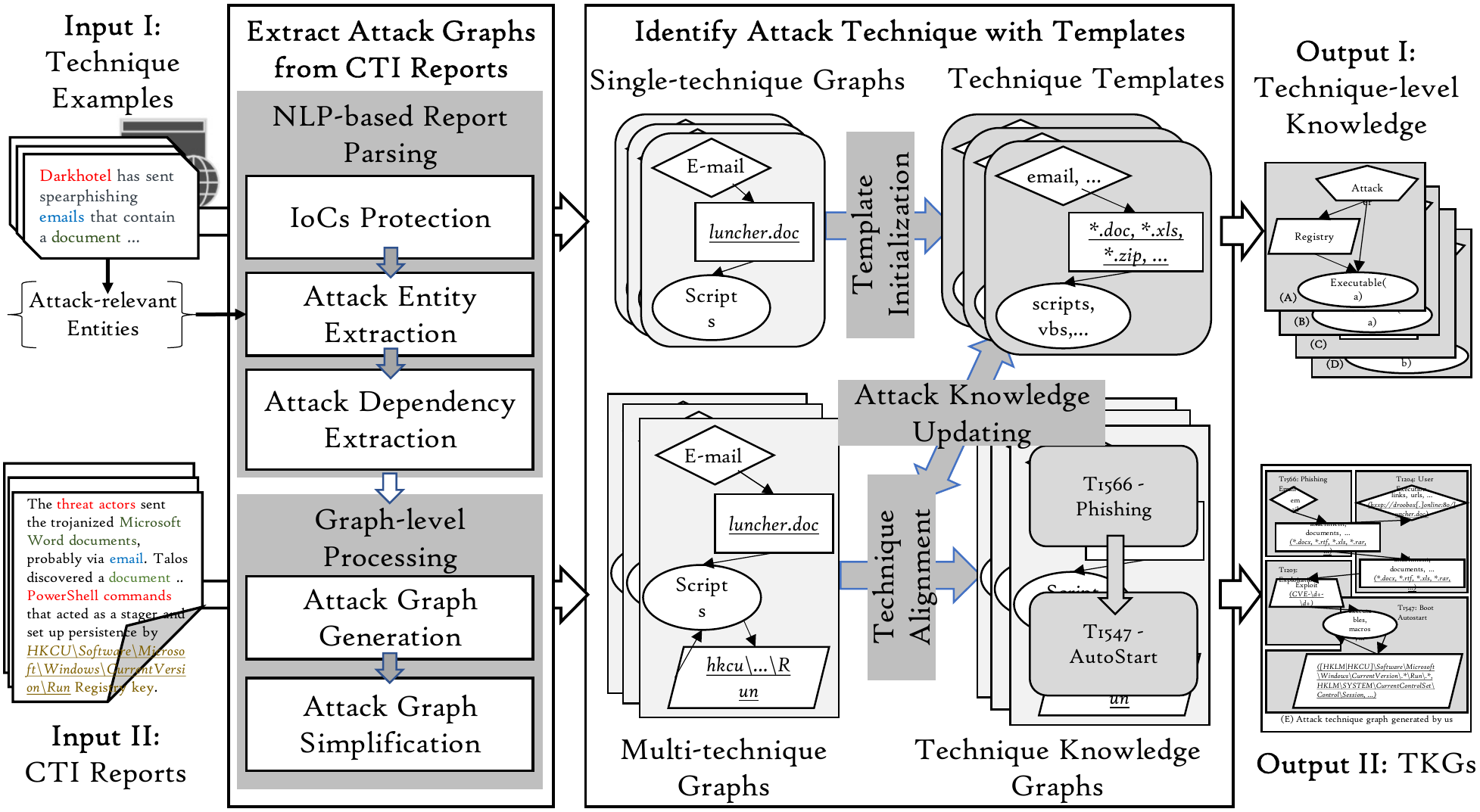}
    \vspace{-0.06in}
    \caption{\small{Overview of \sysname architecture. 
    % \sysname takes two inputs: (1) technique procedure examples from MITRE describing individual attack techniques; (2) real-world CTI reports describing end-to-end attack workflow, and provides two outputs: (1) technique templates that aggregate technique-level threat intelligence across reports; (2) attack graphs enhanced with aggregated technique intelligence. Details of Output I and II can be found in Figure~\ref{fig:template} and~\ref{fig:example}, respectively.
    }}
    \label{fig:architecture}
    \vspace{-0.17in}
\end{figure*}

Figure \ref{fig:architecture} shows the architecture of AttacKG. At a high
level, \sysname has two subsystems: 
(1) an attack graph extraction pipeline for CTI reports parsing and attack graphs building, 
and (2) an attack technique
identification subsystem for technique template generation and technique
identification in attack graphs. 
% We provide a brief overview of the workflow of
% \sysname below, with more implementation details discussed in Section~\ref{sec:parser} and~\ref{sec:templates}, respectively.

\vspace{-0.15in}
\subsubsection{Extract Attack Graphs from CTI Reports.} 
\label{subsec:overview_parser}

To accurately extract attack graphs from CTI reports, we design a parsing pipeline of five stages.
% (1) IoC protection, (2) Attack entities extraction, (3) Attack dependency extraction, (4) Attack graph generation, and (5) Attack graph simplification.
As shown in Figure~\ref{fig:architecture}, this pipeline has two inputs: 
(1) technique procedure examples crawled from MITRE ATT\&CK describing individual techniques, 
and (2) CTI reports describing multi-technique attack campaigns. 
These two inputs are isomorphic, and the corresponding outputs of the pipeline are single-technique graphs for attack techniques and multi-techniques graphs for attack campaigns.

% Specifically, we first extract attack-relevant entities from CTI reports with regular expressions and a customized NLP module. Then, we retrieve intra-sentence dependencies with NLP dependency parsing component and inter-sentence dependencies with resolved co-reference. Finally, we are able to generate the attack graph with entities as nodes and dependencies among them as edges. 

\vspace{-0.15in}
\subsubsection{Identify Attack Technique with Templates.}

As discussed in Section~\ref{sec:background}, individual reports typically have a limited aspect of attack patterns without a global vision. 
In this paper, we aim to bridge this gap by aggregating threat intelligence across CTI reports with technique templates. 
For this purpose, we propose technique templates to aggregate technique-level intelligence and a revised graph alignment algorithm to identify techniques in attack graphs.

As Figure~\ref{fig:architecture} shows, technique templates are initialized with single-technique attack graphs extracted from technique examples crawled from MITRE. 
Then, we adopt the revised attack graph alignment algorithm to identify attack techniques in multi-technique graphs extracted from real-world CTI reports with the pre-initialized templates. 
With the aligned nodes in the attack graph and the corresponding technique template, we can enhance the attack graph with general knowledge in templates into a Technique Knowledge Graph (TKG) and update the technique template with rich intelligence from CTI reports the same time.

Finally, we obtain two outputs: (1) technique templates collecting and aggregating attack knowledge across CTI reports at the technique level; (2) TKGs summarizing complete attack chains in CTI reports.
It is worth mentioning that the template can tolerate a few false-positives/false-negatives in templates or attack graphs as long as techniques implementations appear multiple times in different reports and most of them are parsed correctly.

\subsection{CTI Reports Parser}
\label{sec:parser}

In this section, we introduce our primary approach for extracting attack graphs from CTI reports. Well-written CTI reports include detailed technical descriptions of how attack-related entities interact to accomplish specific objectives in attack campaigns. Despite the rich information, it is challenging to accurately extract attack behaviors from CTI reports written in natural language. Specifically, we identify four main challenges:

\noindent \textbf{[C1] Domain-specific terms identification.} 
CTI reports often contain numerous security-related terms, such as IoCs and attack family names, 
that include special characters and confuse most off-the-shelf NLP modules. 

\noindent \textbf{[C2] Attack entity and dependency extraction.} 
Unlike provenance graphs that record attack actions with full details, 
CTI reports are written in a more summarized manner, 
providing an overview of the attack workflow. 
Thus, the attack graphs extracted from CTI reports only demonstrate coarse-grained and incomplete dependencies among entities.

\noindent \textbf{[C3] Co-reference resolution.} 
Co-reference is very common in natural language. 
We identify two types of co-reference in CTI reports. Explicit co-references use pronouns like ``\textit{it}'' and ``\textit{this}'' or definite article ``\textit{the},''
while implicit co-references use synonyms to refer to entities that appear in the preceding text. 

\noindent \textbf{[C4] Attack graph construction and simplification.} 
Attack scenarios described in natural language are redundant and fractured, which NLP technology cannot address.
Therefore we need to construct and simplify graphs with the assistance of domain knowledge.

To address these challenges, we design a new CTI report parsing pipeline based on the existing ones~\cite{gao,extractor}, with better performance in handling co-reference and constructing attack graphs.
\highlightb{Notice that most of CTI reports are shared in the forms of PDF and HTML, which we further translate them into a uniform text format with open-source tools like \textit{pdfpulmer} and \textit{html2text}.}

\vspace{-0.15in}
\subsubsection{IoC Recognition and Protection with Regex [C1].}
\label{subsec:ioc_protection}

CTI reports contain numerous domain-specific terms, 
such as \texttt{CVE-2017-21880} and \texttt{/etc/passwd/}, 
which include special characters and thus confuse general NLP models. 
In order to avoid the influence of these terms while preserving the information of attack behaviors, 
we identify them with a refined version of open-source IoC recognizer~\cite{iocparser} by extending the regex set and replacing them with commonly used words according to their entity types. 
And we will record the location of replaced words for subsequent resumption of the IoCs.
Then, we are able to adopt standard models~\cite{spacy} for first-stage report parsing. 

\vspace{-0.15in}
\subsubsection{Attack Entity and Dependency Extraction [C2, C3].}
\label{subsection:entities_extraction}

In addition to IoC entities, non-IoC entities also play important roles in attack technique expression. 
For better extraction, 
we classify entities into six types. %, as shown in Table~\ref{tab:entity_types}.
Among them, \textbf{Actor} and \textbf{Executable} represent subjects of attack behaviors, while \textbf{File}, \textbf{Network connection}, and \textbf{Registry} denote common system-level objects. 
Furthermore, we identify several ``\textbf{other}'' types of entities that frequently appear in certain techniques but are difficult to directly map to system objects. 
As a result, we classify them into a separate category named Others.

Then, we adopt a learning-based Named Entity Recognition (NER) model to recognize entities in CTI reports. 
The model is pre-trained on a large general corpus\footnote{\url{https://github.com/explosion/spacy-models/releases/tag/en\_core\_web\_sm-3.1.0}} and re-trained with technique examples randomly sampled from MITRE covering all techniques and entity types. 
To improve the accuracy of entity extraction, 
we further use a customized rule-based entity recognizer\footnote{\url{https://spacy.io/api/entityruler}} to identify well-defined and common entities. 
In addition, we adopt an open-source co-reference resolver, co-referee\footnote{\url{https://github.com/msg-systems/coreferee}}, for explicit co-reference's resolution. 
All pronouns for an attack-relevant entity are recorded in a linked table, and the corresponding nodes will be merged when constructing the attack graph.

For intra-sentence dependency extraction, we first construct a dependency tree for each sentence with a learning-based nature language parsing model~\cite{spacy}. 
Then, we enumerate all pairs of attack-relevant entities (including their pronouns) and estimate the distance between them with the distance of their Lowest Common Ancestor (LCA) and the distance of their position in the sentence. 
Each entity will establish dependencies with its nearest entity unless there exists only one entity in the sentence.

\vspace{-0.15in}
\subsubsection{Attack Graph Generation and Simplification [C3, C4].}
\label{subsec:graph_simplification}

Given extracted attack entities and dependencies, we can initialize a graph with nodes representing attack-relevant entities and edges representing dependencies, which we call \textit{attack graph} ($G_a$) in this paper.
So far, we have only considered the dependencies within sentences. 
Cross-sentence dependencies will be established through both explicit and implicit co-reference nodes.
Specifically, by merging co-reference nodes, we not only remove redundant nodes but also combine sentence-level sub-graphs into a whole attack graph.
Explicit co-reference can be recognized by general NLP models, as discussed in Section \ref{subsection:entities_extraction}, 
while implicit co-references need to be identified based on entities' type and character-level overlaps. 
In particular, we use node-level alignment scores, as discussed in Section~\ref{subsec:TT_Identification}, to determine whether two nodes should be treated as co-references nodes.

Finally, we generate a concise and clear attack graph describing all attack behaviors that appear in a CTI report.
The evaluation on a wide range of CTI reports demonstrates that our attack graph extraction pipeline is both accurate and effective (Section~\ref{subsec:rq3}).
And a few false-positives/false-negatives have limited impact on the subsequent fuzzy alignment-based technique identification.

\vspace{-0.15in}
\subsection{Technique Templates and Graph Alignment Algorithm}
\label{sec:templates}
\vspace{-0.06in}

In order to identify specific techniques from the attack graph while accurately extracting the corresponding threat intelligence mentioned in the CTI report, we first need a universal description of the attack technique. In this paper, we adopt a design of technique template in a graph structure for each technique.

Then, inspired by the graph alignment algorithm proposed in Poirot~\cite{Milajerdi2019} for attack behavior identification in provenance graphs, we design a revised graph alignment algorithm for technique identification with templates.
Finally, we introduce how to initialize and update technique templates with alignment results.

\vspace{-0.15in}
\subsubsection{Design of Technique Templates}
\label{subsec:TT_Design}

To present the attack behaviors inside techniques while aggregating threat intelligence, we model technique templates also as graphs ($G_t$) with statistics information. 
In the graph, nodes represent aggregated entity knowledge, and edges represent possible dependencies among them. 

% https://www.tablesgenerator.com/latex_tables
\begin{table*}[ht]    
\footnotesize
\centering
\vspace{-0.2in}
\caption{Attack entities in the template of the technique \texttt{T1547-Boot or Logon Autostart Execution.}}
\begin{adjustbox}{width=\textwidth}
\begin{tabular}{cll}
\toprule
\textbf{\makecell{Template\\ entities}}    & \textbf{\makecell{NLP\\ descriptions}}                                                    & \textbf{IoC   terms}                                                                                                                                                                                                                                                                                              \\ \midrule
Executable       & \makecell{scripts,\\macros,  …  }                                                              & *.exe,   *.ps1, …                                                                                                                                                                                                                                                                                                 \\ \hline
Register         & \makecell{register keys, \\ register, …  }                                                               & \makecell[l]{ HKLM\textbackslash{}…\textbackslash{}windows\textbackslash{}currentversion\textbackslash{}winlogon\textbackslash{}*,\\  HKLM\textbackslash{}…\textbackslash{}active setup\textbackslash{}installed components\textbackslash{}*, \\HKLM\textbackslash{}Software\textbackslash{}*\textbackslash{}Run, …} \\ \hline
Autostart Folder & \makecell{ startup folder,\\ path, … }                                                                 & \makecell[l]{ \%HOMEPATH\%\textbackslash{}Start Menu\textbackslash{}Programs\textbackslash{}Startup\textbackslash, \\$\sim$/.config/autostart/*, …}                                                                                              \\ \hline
Shortcut File  & \makecell{shortcut, …  }                                                                               & *.lnk, …                                                                                                                                                                                                                                                                                                       \\ \hline
Side-loading DLL              & \makecell{winlogon helper DLL,\\ SSP DLL, …} & \begin{tabular}[c]{@{}l@{}} sspisrv.dll, …\end{tabular}                                                                                                                                                                                                                                                         \\ \bottomrule
\end{tabular}
\end{adjustbox}
\label{tab:template}
\vspace{-0.2in}
\end{table*}

% \begin{figure*}[ht]
%     \centering
%     \includegraphics[width=1\textwidth]{Image/Template.pdf}
%     \vspace{-0.15in}
%     \caption{\highlightb{\small{Technique template generated by \sysname for \texttt{T1547 - Boot or Logon Autostart Execution} with four variants corresponding to fourteen MITRE sub-techniques categorized as (A) Registry Run Keys, (B) Auto-start folder, (C) Shortcut Modification, and (D) DLL Side-loading.}}}
%     \label{fig:template}
% \vspace{-0.15in}
% \end{figure*}

% \highlightb{Our observation is that the cyber attacks have lots of variants, but most of them are build from some template.}

Moreover, we designed a confidence score calculated by the number of occurrences for entities and dependencies in different reports. In this way, as long as techniques appear multiple times in different reports and most of them are parsed correctly, the impact of possible false-positives/false-negatives (FPs/FNs) involved by AttacKG misidentifications and adversarial or low-quality CTI reports can be tolerated. 

\vspace{-0.15in}
\subsubsection{Graph Alignment for Technique Identification and Technique Knowledge Graph Construction.}
\label{subsec:TT_Identification}

As discussed above, both attack graphs we extracted in Section~\ref{sec:parser} and technique templates we generated in Section~\ref{sec:templates} may contain false-positives/false-negatives (FPs/FNs), so we cannot use exact match. As an alternative, we proposed a graph alignment algorithm between technique template $G_t$ and attack graph $G_a$ for fuzzy matching.
Specifically, as shown in Table~\ref{tab:notation}, we define two kinds of alignments, i.e., node alignment between two nodes in two different graphs and graph alignment measuring the overall similarity between a technique template and a certain subgraph in the attack graph. 

\begin{table}[htbp]
\centering
\vspace{-0.15in}
\caption{Notations in Graph Alignment}
\label{tab:notation}
\begin{adjustbox}{width=\textwidth}
\begin{tabular}{@{}cl@{}}
\toprule
Notation             & Description                                                   \\ \midrule
$i : k$              & Node alignment between node $i$ and $k$ from different graph           \\
% $i \dashrightarrow  j$   & An edge from node $i$ to node $j$                   \\
$i \dashrightarrow  j$   & An dependency (path) from node $i$ to node $j$                   \\
$G_t :: G_a$         & Graph alignment between Template $G_t$ and Graph $G_a$ \\ 
$\Gamma(i : k)$      & Alignment score between node $i$ to node $k$                  \\
$\Gamma(G_t :: G_a)$      & \makecell[l]{Alignment score between Template $G_t$ and Graph $G_a$}   \\
$\Gamma_N(G_t :: G_a)$      & \makecell[l]{Node-level alignment score between Template $G_t$ and Graph $G_a$}   \\
$\Gamma_E(G_t :: G_a)$      & \makecell[l]{Dependency-level alignment score between Template $G_t$ and Graph $G_a$}    \\ \bottomrule
\end{tabular}
\end{adjustbox}
\vspace{-0.15in}
\end{table}

\vspace{0.03in}
\textbf{Node alignment.} We first enumerate every node $k$ in attack graph $G_a$ to find alignment candidates for every node $i$ in a template $G_t$ by calculating the alignment score for nodes $\Gamma(i : k)$. 
The alignment score between two nodes is computed by Equations (\ref{eq:nodealign}) and (\ref{eq:sim}):

{
\small
\vspace{-0.09in}
\begin{equation}
\label{eq:nodealign}
\Gamma(i : k)=
\begin{cases}
\gamma + (1-\gamma) \cdot Sim(i, k) & i_{type} = k_{type} \\ 
0 &  i_{type} \neq k_{type} 
\end{cases},
\end{equation}

\vspace{-0.09in}
\begin{equation}
\label{eq:sim}
Sim(i, k) = Max(sim(i_{IoC}, k_{IoC}), sim(i_{NLP}, k_{NLP})).
\end{equation}
\vspace{-0.09in}
}

Intuitively, if node $i$ and node $k$ have different types, then the alignment score will be zero. Otherwise, they will get a basic type-matched score $\gamma$. Then the similarity of nodes' attributes ($Sim(i, k)$) can be determined by enumerating and calculating terms in the IoC term set and natural language description set similarity ($Sim(i, k)$) at the character level~\cite{levenshtein}. If the alignment score reaches a pre-defined threshold, we will record the node alignment candidate in $G_a$ to a list of the corresponding template node. 

\vspace{0.03in}
\textbf{Graph alignment.} Afterward, we iterate through all candidate nodes to calculate the overall alignment scores $\Gamma(G_t :: G_a)$ in two parts: node-level alignment scores $\Gamma_N(G_t :: G_a)$ and edge-level alignment scores $\Gamma_E(G_t :: G_a)$.
Specifically, the alignment score between a technique template and an attack graph can be computed by Equations (\ref{eq:AN}), (\ref{eq:AD}), and (\ref{eq:AO}):

{
\small

\vspace{-0.09in}
\begin{equation}
\label{eq:AN}
   \begin{aligned}
    \Gamma_N(G_t :: G_a) = \sum_{i \in G_t, k \in G_a}(\Gamma(i : k) \cdot i_{occur}) \Big{/} \sum_{i \in G_t}(i_{occur})
    \end{aligned},
\end{equation}

\vspace{-0.09in}
\begin{equation}
\label{eq:AD}
   \begin{aligned}
    \Gamma_E(G_t :: G_a) =& \sum_{\footnotesize{\makecell{i \dashrightarrow j \in G_t\\k \dashrightarrow l \in G_a}}} (\frac{(\Gamma(i : k) \cdot \Gamma(j : l)}{C_{min}(k \dashrightarrow l)} \cdot (i \dashrightarrow j)_{occur})\\
   & \Big{/} \sum_{i \dashrightarrow j \in G_a} ((i \dashrightarrow j)_{occur}),
    \end{aligned}
\end{equation}

\vspace{-0.09in}
\begin{equation}
\label{eq:AO}
    \Gamma(G_t :: G_a) = \frac{1}{2} \cdot (\Gamma_N(G_t :: G_a) + \Gamma_E(G_t :: G_a)).
\end{equation}
\vspace{-0.09in}
}

As shown, the node-level alignment score ($\Gamma_N(G_t :: G_a)$) is a weighted sum of the alignment score ($\Gamma(i : k)$) of each node. The weights are proportional to the number of node occurrences ($i_{occur}$) recorded in the template.
In this way, we enhance the impact of important entities and dependencies that commonly appear in different reports.
Meanwhile, the edge-level alignment score ($\Gamma_E(G_t :: G_a)$) depends on three factors: 
alignment scores of the nodes at both ends of the dependency ($\Gamma(i : k)$ and $\Gamma(j : l)$); 
the minimal hop between both ends of the dependency ($C_{min}(k \dashrightarrow l)$) in the attack graph; 
the number of node occurrences ($(i \dashrightarrow j)_{occur}$) recorded in the template. 
If two nodes are not connected, the dependency between them ($C_{min}(k \dashrightarrow l)$) will be set to infinity. 
Finally, the outputs of all five equations above are normalized to interval $[0,1]$.
We note that although the traversal method is used, the computational overhead of the whole algorithm is acceptable due to the limited size of both attack graph and technique templates.

After getting alignment scores for candidate permutations of each technique, we will compare them with a pre-defined threshold and finally select aligned subgraphs of techniques. It is noteworthy that one attack graph node can be aligned in multiple techniques, and one technique can be found multiple time as long as each aligned subgraph have an alignment score above the threshold.

\vspace{0.03in}
\textbf{TKG construction.} With the graph alignment results, we can attach the general knowledge stored in technique templates, including alternative entities and techniques, to the corresponding positions in an attack graph. Then, we can obtain the Technique Knowledge Graph (TKG) that introduces the whole attack chain in a CTI report with enhanced knowledge.

\vspace{-0.15in}
\subsubsection{Initialization and Updating of Technique Templates }
\label{subsec:TT_generation}

Both the initialization and updating of technique templates rely on the graph alignment results. 

\vspace{0.03in}
\textbf{Updating.} 
By aligning, the node in the attack graph can be mapped to the node in the identified technique template. 
Then we update the IoC and natural language description sets in the template node with new terms from the aligned attack graph node. 
This allows us to aggregate threat intelligence across CTI reports at the technique level. 

\vspace{0.03in}
\textbf{Initialization.} 
The initialization process of technique templates starts with a random single-technique attack graph extracted from MITRE technique examples as the initial template. 
Then we align the initial template with other single-technique attack graphs of the same technique. 
The information in aligned attack graph nodes will be merged into the corresponding template node. And the unaligned nodes will be added to templates as new nodes, which is different from the updating process. 

Finally, we generate the technique template that aggregates threat intelligence and covers multiple technique variants across multiple reports, as the example shown in Figure~\ref{fig:template} and Table~\ref{tab:template}.

\section{Evaluation}
\label{sec:evaluation}
\vspace{-0.06in}

In this section, we focus on evaluating AttacKG's accuracy of attack graph extraction and technique identification as a CTI report parser and its effectiveness of technique-level intelligence aggregation as a CTI knowledge collector.
In particular, our evaluation aims at answering the following research questions (RQs): (\textbf{RQ1}) How accurate is \sysname in extracting attack graphs (attack-related entities and dependencies) from CTI reports? (\textbf{RQ2}) How accurate is \sysname in identifying attack techniques in CTI reports? (\textbf{RQ3}) How effective is \sysname in collecting technique-level intelligence from massive CTI reports? Finally, we also want to evaluate how Technique Knowledge Graphs benefit downstream security tasks.

\subsection{Evaluation Setup}
\label{subsec:eval_setup}
\highlightb{To evaluate AttacKG, we crawled 1,515 real-world CTI reports mentioned in MITRE ATT\&CK references whose sources range from Cisco Talos Intelligence Group~\cite{ciscotalos}, Microsoft Security Intelligence Center~\cite{microsoft}, etc.}
Moreover, we crawl 7,373 procedure examples out of 179 techniques from the MITRE ATT\&CK knowledge-base~\cite{mitre} to formulate our technique templates.

To answer RQ1 and RQ2, we further manually label the ground-truth of entities, dependencies, and techniques in 16 of the collected reports:
(1) \textit{Five DARPA TC Reports}:
We select five attack reports released by the DARPA TC program's fifth engagement that cover different OS platforms (i.e., Linux, Windows, and FreeBSD), vulnerabilities (e.g., Firefox backdoor), and exploits (e.g., Firefox BITS Micro).
(2) \textit{Three APT Campaign Reports}: 
To explore the performance of \sysname in practice, we select another three public CTI reports that describe APT campaigns from three well-known threat groups, i.e., Frankenstein~\cite{frankenstein}, OceanLotus (APT32)~\cite{oceanlotus}, and Cobalt Group~\cite{cobalt}. 

\subsection{Evaluation Results}

\subsubsection{RQ1: How accurate is \sysname in extracting attack graphs from CTI reports?} 
\label{subsec:rq1}

A typical attack technique consists of multiple threat actions that are presented as a set of connected entities in an attack graph.
In particular, the accurate extraction of attack graphs is an essential starting point towards automated identification of attack techniques from CTI reports.
To evaluate the accuracy of \sysname in extracting attack graphs, we adopt the aforementioned 16 well-labeled CTI reports.
We manually identify attack-related entities in the reports and correlate entities based on our domain knowledge of the attack workflow.
It is noteworthy that in addition to natural language descriptions, DARPA TC reports also provide the graph representation of attacks, which serves as additional documentation to complement our manual labels.
% A detailed labeling result report example is available in Appendix~\ref{appendix:casestudy}.

Given ground-truth entities and dependencies in the reports, we are able to compare \sysname with the state-of-the-art open-source\footnote{\url{https://github.com/ksatvat/EXTRACTOR}} CTI report parser, EXTRACTOR~\cite{Satvat2021}, in terms of the precision, recall, and F1-score. 
For a fair comparison, we enable all optimizations in EXTRACTOR (e.g., Ellipsis Subject Resolution) when constructing attack graphs upon textual attack descriptions.
As discussed in Section~\ref{sec:system_overview}, an entity may correspond to multiple co-references across a CTI report.
Since our goal is to identify unique entities (e.g., IOCs), we merge co-reference entities in the attack graph and integrate their dependencies with the remaining entities.

\highlightb{
\begin{table}[]
    \centering
    \scriptsize
    \vspace{-0.15in}
    \caption{\highlightb{\small{Accuracy of attack graph extraction and technique identification in 16 CTI reports. (Columns 2-9 present the count of manually-generated ground-truth and $\textcolor{fncolor}{-false\_negative} (\textcolor{fpcolor}{+false\_positive})$ in extracting attack-related entities, dependencies, and techniques. Columns 10-12 present the overall Precision, Recall and F1-score.)}}}
    \label{tab:threat_behavior_accuracy}
    \begin{adjustbox}{width=\textwidth}
    \begin{tabular}{l|ccc|ccc|ccc}
    \toprule
    \multirow{2}{*}{CTI reports}              & \multicolumn{3}{c}{Entities}                       & \multicolumn{3}{c}{Dependencies}                       & \multicolumn{3}{c}{Techniques}                  \\ \cmidrule(l){2-10} 
                                            & Manual         & Extractor      & \sysname           & Manual         & Extractor      & \sysname           & Manual         & TTPDrill       & \sysname           \\ \midrule
    TC\_Firefox DNS Drakon APT              & 10             & \textcolor{fncolor}{-4} (\textcolor{fpcolor}{+4})         & \textcolor{fncolor}{-0} (\textcolor{fpcolor}{+1})         & 9              & \textcolor{fncolor}{-4} (\textcolor{fpcolor}{+3})         & \textcolor{fncolor}{-2} (\textcolor{fpcolor}{+1})         & 8              & \textcolor{fncolor}{-2} (\textcolor{fpcolor}{+10})        & \textcolor{fncolor}{-0} (\textcolor{fpcolor}{+3})         \\ 
    TC\_Firefox Drakon Copykatz & 6              & \textcolor{fncolor}{-2} (\textcolor{fpcolor}{+0})         & \textcolor{fncolor}{-1} (\textcolor{fpcolor}{+0})         & 5              & \textcolor{fncolor}{-2} (\textcolor{fpcolor}{+0})         & \textcolor{fncolor}{-2} (\textcolor{fpcolor}{+0})         & 4              & \textcolor{fncolor}{-1} (\textcolor{fpcolor}{+13})        & \textcolor{fncolor}{-1} (\textcolor{fpcolor}{+0})         \\ 
    TC\_Firefox BITS Micro APT              & 11             & \textcolor{fncolor}{-6} (\textcolor{fpcolor}{+0})         & \textcolor{fncolor}{-1} (\textcolor{fpcolor}{+4})         & 10             & \textcolor{fncolor}{-7} (\textcolor{fpcolor}{+0})         & \textcolor{fncolor}{-0} (\textcolor{fpcolor}{+0})         & 5              & \textcolor{fncolor}{-1} (\textcolor{fpcolor}{+14})        & \textcolor{fncolor}{-2} (\textcolor{fpcolor}{+2})         \\ 
    TC\_SSH BinFmt-Elevate                  & 6              & \textcolor{fncolor}{-4} (\textcolor{fpcolor}{+0})         & \textcolor{fncolor}{-1} (\textcolor{fpcolor}{+0})         & 5              & \textcolor{fncolor}{-4} (\textcolor{fpcolor}{+0})         & \textcolor{fncolor}{-0} (\textcolor{fpcolor}{+0})         & 5              & \textcolor{fncolor}{-2} (\textcolor{fpcolor}{+14})        & \textcolor{fncolor}{-2} (\textcolor{fpcolor}{+2})         \\ 
    TC\_Nginx Drakon APT                    & 15             & \textcolor{fncolor}{-2} (\textcolor{fpcolor}{+0})         & \textcolor{fncolor}{-2} (\textcolor{fpcolor}{+0})         & 15             & \textcolor{fncolor}{-0} (\textcolor{fpcolor}{+0})         & \textcolor{fncolor}{-2} (\textcolor{fpcolor}{+0})         & 6              & \textcolor{fncolor}{-2} (\textcolor{fpcolor}{+22})        & \textcolor{fncolor}{-0} (\textcolor{fpcolor}{+2})         \\ 
    Frankenstein Campaign                   & 14             & \textcolor{fncolor}{-3} (\textcolor{fpcolor}{+1})         & \textcolor{fncolor}{-0} (\textcolor{fpcolor}{+2})         & 16             & \textcolor{fncolor}{-5} (\textcolor{fpcolor}{+1})         & \textcolor{fncolor}{-0} (\textcolor{fpcolor}{+2})         & 9              & \textcolor{fncolor}{-1} (\textcolor{fpcolor}{+18})        & \textcolor{fncolor}{-1} (\textcolor{fpcolor}{+1})         \\ 
    OceanLotus(APT32) Campaign              & 7              & \textcolor{fncolor}{-0} (\textcolor{fpcolor}{+2})         & \textcolor{fncolor}{-0} (\textcolor{fpcolor}{+2})         & 7              & \textcolor{fncolor}{-0} (\textcolor{fpcolor}{+1})         & \textcolor{fncolor}{-1} (\textcolor{fpcolor}{+0})         & 5              & \textcolor{fncolor}{-1} (\textcolor{fpcolor}{+12})        & \textcolor{fncolor}{-2} (\textcolor{fpcolor}{+0})         \\ 
    Cobalt Campaign                         & 17             & \textcolor{fncolor}{-6} (\textcolor{fpcolor}{+0})         & \textcolor{fncolor}{-1} (\textcolor{fpcolor}{+5})         & 17             & \textcolor{fncolor}{-4} (\textcolor{fpcolor}{+0})         & \textcolor{fncolor}{-1} (\textcolor{fpcolor}{+2})         & 8              & \textcolor{fncolor}{-2} (\textcolor{fpcolor}{+21})        & \textcolor{fncolor}{-1} (\textcolor{fpcolor}{+1})         \\ 
    DeputyDog Campaign                         & 13             & \textcolor{fncolor}{-1} (\textcolor{fpcolor}{+2})         & \textcolor{fncolor}{-0} (\textcolor{fpcolor}{+2})         & 14             & \textcolor{fncolor}{-1} (\textcolor{fpcolor}{+1})         & \textcolor{fncolor}{-2} (\textcolor{fpcolor}{+0})         & 10              & \textcolor{fncolor}{-1} (\textcolor{fpcolor}{+35})        & \textcolor{fncolor}{-0} (\textcolor{fpcolor}{+6})         \\ 
    HawkEye Campaign                         & 16             & \textcolor{fncolor}{-2} (\textcolor{fpcolor}{+3})         & \textcolor{fncolor}{-3} (\textcolor{fpcolor}{+4})         & 17             & \textcolor{fncolor}{-5} (\textcolor{fpcolor}{+3})         & \textcolor{fncolor}{-3} (\textcolor{fpcolor}{+2})         & 11              & \textcolor{fncolor}{-2} (\textcolor{fpcolor}{+64})        & \textcolor{fncolor}{-1} (\textcolor{fpcolor}{+3})         \\ 
    DustySky Campaign                         & 12             & \textcolor{fncolor}{-2} (\textcolor{fpcolor}{+1})         & \textcolor{fncolor}{-0} (\textcolor{fpcolor}{+3})         & 12             & \textcolor{fncolor}{-2} (\textcolor{fpcolor}{+1})         & \textcolor{fncolor}{-0} (\textcolor{fpcolor}{+3})         & 5              & \textcolor{fncolor}{-0} (\textcolor{fpcolor}{+32})        & \textcolor{fncolor}{-0} (\textcolor{fpcolor}{+1})         \\ 
    TrickLoad Spyware Campaign                         & 17             & \textcolor{fncolor}{-3} (\textcolor{fpcolor}{+1})         & \textcolor{fncolor}{-0} (\textcolor{fpcolor}{+0})         & 16             & \textcolor{fncolor}{-4} (\textcolor{fpcolor}{+0})         & \textcolor{fncolor}{-0} (\textcolor{fpcolor}{+1})         & 4              & \textcolor{fncolor}{-0} (\textcolor{fpcolor}{+18})        & \textcolor{fncolor}{-2} (\textcolor{fpcolor}{+0})         \\ 
    Emotet Campaign                         & 8             & \textcolor{fncolor}{-4} (\textcolor{fpcolor}{+0})         & \textcolor{fncolor}{-1} (\textcolor{fpcolor}{+1})         & 7             & \textcolor{fncolor}{-4} (\textcolor{fpcolor}{+0})         & \textcolor{fncolor}{-2} (\textcolor{fpcolor}{+1})         & 7              & \textcolor{fncolor}{-2} (\textcolor{fpcolor}{+16})        & \textcolor{fncolor}{-3} (\textcolor{fpcolor}{+0})         \\ 
    Uroburos  Campaign                         & 12             & \textcolor{fncolor}{-1} (\textcolor{fpcolor}{+2})         & \textcolor{fncolor}{-2} (\textcolor{fpcolor}{+3})         & 13             & \textcolor{fncolor}{-3} (\textcolor{fpcolor}{+0})         & \textcolor{fncolor}{-2} (\textcolor{fpcolor}{+0})         & 7              & \textcolor{fncolor}{-0} (\textcolor{fpcolor}{+23})        & \textcolor{fncolor}{-1} (\textcolor{fpcolor}{+2})         \\ 
    APT41  Campaign                         & 13             & \textcolor{fncolor}{-1} (\textcolor{fpcolor}{+5})         & \textcolor{fncolor}{-1} (\textcolor{fpcolor}{+0})         & 12             & \textcolor{fncolor}{-0} (\textcolor{fpcolor}{+1})         & \textcolor{fncolor}{-1} (\textcolor{fpcolor}{+2})         & 6              & \textcolor{fncolor}{-2} (\textcolor{fpcolor}{+26})        & \textcolor{fncolor}{-1} (\textcolor{fpcolor}{+1})         \\ 
    Espionage  Campaign                         & 11             & \textcolor{fncolor}{-2} (\textcolor{fpcolor}{+6})         & \textcolor{fncolor}{-3} (\textcolor{fpcolor}{+1})         & 10             & \textcolor{fncolor}{-3} (\textcolor{fpcolor}{+2})         & \textcolor{fncolor}{-3} (\textcolor{fpcolor}{+1})         & 4              & \textcolor{fncolor}{-0} (\textcolor{fpcolor}{+19})        & \textcolor{fncolor}{-0} (\textcolor{fpcolor}{+1})         \\ \midrule
    \textbf{Overall Presicion}              & \textbf{1.000} & \textbf{0.843} & \textbf{0.860} & \textbf{1.000} & \textbf{0.913} & \textbf{0.906} & \textbf{1.000} & \textbf{0.196} & \textbf{0.771} \\ 
    \textbf{Overall Recall}                 & \textbf{1.000} & \textbf{0.771} & \textbf{0.915} & \textbf{1.000} & \textbf{0.741} & \textbf{0.886} & \textbf{1.000} & \textbf{0.837} & \textbf{0.808} \\ 
    \textbf{Overall F-1 Score}              & \textbf{1.000} & \textbf{0.806} & \textbf{0.887} & \textbf{1.000} & \textbf{0.818} & \textbf{0.896} & \textbf{1.000} & \textbf{0.318} & \textbf{0.789} \\ \bottomrule
    \end{tabular}
    \end{adjustbox}
    \vspace{-0.15in}
\end{table}
}

Table~\ref{tab:threat_behavior_accuracy} summarizes the results of \sysname and EXTRACTOR in capturing entities and dependencies from the selected 16 CTI reports(Rows 2-7). 
As can be seen, despite sightly lower precision caused by a higher false-positive rate, \sysname yields better accuracy overall (with an average F1-score improvement of 0.12) than EXTRACTOR due to a much lower false-negative rate.
This is expected as EXTRACTOR aggregates all non-IoC entities of the same type (e.g., process) into one entity, as shown in Figure~\ref{fig:example}.
In other words, no matter how many false-positive entities EXTRACTOR produces, they are treated as one false extraction as long as they belong to the same type.
\highlightb{It is noteworthy that such aggregation design inevitably losses structural information of attack graphs and makes follow-up technique identification almost impossible.}
Hence, we only compare \sysname with EXTRACTOR in extracting attack graphs rather than identifying attack techniques. 

\vspace{-0.1in}
\subsubsection{RQ2: How accurate is \sysname in identifying attack techniques in CTI reports?}
\label{subsec:rq2}

To answer RQ2, we use \sysname to identify attack techniques in the 16 CTI reports and compare it with the state-of-the-art technique identifier, TTPDrill~\cite{Husari2017}.
The core idea of TTPDrill is to extract threat actions from CTI reports and attribute such actions to techniques based on threat-action ontology.
Specifically, it manually defines 392 threat actions for 187 attack techniques in the original paper, while such ontology knowledge base has been extended to cover 3,092 threat actions for 246 attack techniques in its latest open-source implementation\footnote{\url{https://github.com/mpurba1/TTPDrill-0.3}}.
Also noteworthy is that all attack techniques used by TTPDrill are derived from an old version of MITRE ATT\&CK matrix.
To allow for a consistent comparison, we map every technique in TTPDrill to the latest version technique via the hyperlinks provided by MITRE.
For example, \textit{T1086-PowerShell} in TTPDrill is updated to \textit{T1059/001-Command and Scripting Interpreter: PowerShell}.
% \footnote{\url{https://attack.mitre.org/techniques/T1086}}
% \footnote{\url{https://attack.mitre.org/techniques/T1059/001}}

We evaluate \sysname and TTPDrill on the 16 CTI reports that we labelled the precious ground-truth techniques adopted in the attacks.
The technique identification results are summarized in the last three rows in Table~\ref{tab:threat_behavior_accuracy}.
We can observe that while both \sysname and TTPDrill achieve reasonably low false-negative rates, TTPDrill is prone to high volumes of false-positive techniques (15.5 false positives per report on average), which is nearly three times as many as the true positives.
As a result, while the recall of \sysname is only slightly higher than TTPDrill by 0.1, \sysname significantly outperforms TTPDrill in terms of the precision and F1-score by \highlightb{0.575 and 0.462}, respectively.
This result makes sense as TTPDrill treats threat actions extracted from CTI reports as action-object pairs.
Accordingly, techniques share partial threat actions tend to look similar to each other in TTPDrill.
In contrast, \sysname aligns techniques to attack graphs, taking into consideration the full contexts of threat actions.

Moreover, it is worth mentioning that we use fuzzy matching based on alignment scores for technique identification; thus, our approach can correctly identify attack techniques even with FPs/FNs in technique templates and extracted attack graphs. 
Our observation is that the overall accuracy is highest when the graph alignment score's threshold is 0.85,
% , which means we can tolerate approximately 15\% of errors in all entities and dependencies. 
\highlightb{with details about the threshold selection discussed in Appendix~\ref{sec:threshold}. 
To verify the importance of each component in \sysname towards technique identification, we perform an ablation study by considering four variants of AttacKG, as discussed in Appendix~\ref{sec:ablation_study}.}

\vspace{-0.1in}
\subsubsection{RQ3: How effective is \sysname at collecting technique-level intelligence from massive reports?}
\label{subsec:rq3}

To answer RQ3, we explore the effectiveness of \sysname in extracting threat intelligence (e.g., techniques and IoCs entities) on 1,515 CTI reports collected from different intelligence sources.
Table~\ref{tab:effectiveness} lists the ten most common techniques that appeared in the 1,515 reports and the number of their corresponding unique IoCs,
which mostly overlap with manually generated top TTP lists by PICUS~\cite{picus} and redcanary~\cite{redcanary}.

\begin{table*}[]
\small
\centering
\vspace{-0.25in}
\caption{Effectiveness of Threat Intelligence Extraction from 1,515 CTI Reports.}
% \vspace{-0.04in}
\label{tab:effectiveness}
\begin{adjustbox}{width=\textwidth}
\begin{tabular}{lcccccc}
\toprule
\multirow{2}{*}{Top 6 Techniques}               & \multirow{2}{*}{\begin{tabular}[c]{@{}c@{}}Occurrences \\      in reports\end{tabular}} & \multicolumn{5}{c}{\#Unique IoCs}                                            \\ \cline{3-7} 
                                          &                                                                                         & Executable   & Network       & File & Registry     & Vulner. \\ \midrule
T1071 - Command \& Control                & 1113                                                                                    & 12           & 452           & 371                 & -            & 12            \\ 
T1059 - Scripting Interpreter & 1089                                                                                    & 6            & 394           & 284                 & 100          & 9             \\
T1083 - File/Directory Discovery      & 1060                                                                                    & -            & -             & 249                 & -            & -             \\ 
T1170 - Indicator Removal          & 990                                                                                     & 6            & -             & 255                 & 74           & 7             \\ 
T1105 - Ingress Tool Transfer             & 990                                                                                     & -            & 389           & 261                 & -            & -             \\ 
T1003 - OS Credential Dumping             & 961                                                                                     & -            & -             & 220                 & -            & -             \\ 
% T1204 - User Execution                    & 862                                                                                     & -            & 209           & 180                 & -            & -             \\ 
% T1566 - Phishing                          & 839                                                                                     & 6            & 267           & 307                 & -            & 5             \\ 
% T1574 - Hijack Execution Flow             & 816                                                                                     & -            & -             & 70                  & -            & -             \\ 
% T1005 - Data from Local System            & 792                                                                                     & -            & -             & 197                 & -            & -             \\ 
% T1218 - Signed Binary Proxy Execution     & 664                                                                                     & 17           & -             & 81                  & 37           & -             \\
 \midrule
\textbf{All Techniques Summary}           & \textbf{28262}                                                                          & \textbf{495} & \textbf{2813} & \textbf{4634}       & \textbf{384} & \textbf{67}   \\ \bottomrule
\end{tabular}
\end{adjustbox}
\vspace{-0.2in}
\end{table*}

Each report, on average, contains 18.7 techniques and 5.5 unique IoCs, and different techniques in most cases involve different IoCs.
Most CTI reports do not provide unified and formatted intelligence to validate our extracted results, which is also our motivation behind this work.
Therefore, we randomly select several technique templates with aggregated knowledge for manual investigation. 
Specifically, we observe that templates successfully collect unique IoCs for different technique implementations across CTI reports.
As the example in Figure~\ref{fig:template} and Table~\ref{tab:template} shows, we identify multiple unique IoC terms playing similar roles from different reports for individual implementations (sub-techniques).
Such template-aggregated intelligence can directly enrich our understanding of attack techniques. 
Furthermore, the TKG built on templates can help understand the entire attack and possible variants for more robust detection and investigation, as shown in Section~\ref{subsec:motivating} and Appendix~\ref{appendix:casestudy}.

\highlightb{
\subsection{Case Study}

This subsection discusses how TKGs can be adopted in real-world security tasks with case studies.
Specifically, TKGs adopt the collated knowledge to enrich reports, thus helping to understand and reconstruct the attacks involved. In addition, TKGs with aggregated technique-level intelligence can enhance the detection of attack variants (Appendix~\ref{appendix:casestudy}).
%, namely, ``TKG for attack variants detection'' and ``TKG for attack reconstruction''(discussed in Appendix~\ref{appendix:casestudy}).

%As the motivating example shown in Figure~\ref{fig:example} (E), by constructing attack graphs from CTI reports and align them with attack technique templates, TKGs provide rich behavior information for analysts including attack patterns, insights of vulnerabilities and their associated attack techniques. 
%While CTI largely provides high-level information,
%fine-grained technical details that are relevant for analysis are generally omitted. 

\textbf{TKG for attack reconstruction.} In order for security practitioners and researchers to have an in-depth analysis, they have to bridge the knowledge gap between the real attacks and the CTI reports. 
This gap can be addressed by having a first-hand practical environment that thoroughly describes how attack steps are performed in the CTI reports. 
\sysname provides structured knowledge about an attack scenario, making it easier to reproduce cyber attacks in a testbed environment, benefiting analysts~\cite{uetz2021reproducible} with high fidelity and live reconstructed environment with in-depth details. 
We used \sysname to support attack reconstruction~\cite{liposter}. 

Taking the Frankenstein campaign as an example, with the TKG extracted
from the corresponding report, we can quickly identify nine techniques for six tactics involved in the campaign, including \textit{T1566-Phishing} for tactic Initial Access, \textit{T1547-Boot Autostart} for tactic Persistence, \textit{T1203-Exploitation for Execution} for tactic Execution, etc.
Then, we can infer the environment needed to reconstruct the attack based on the techniques and entities involved in the attack.
Specifically, autostart with registry hints that the attack is running in Windows. The use of vulnerability (\textit{CVE-2017-11882}) for execution indicates a requirement for specific versions of Microsoft Office.
After setup the environment, we can reproduce the campaign with open-source attack technique implementation, such as Atomi-Red-Teams~\cite{atomic}.
All in all, AttacKG provides much necessary information as the first step in the reconstruction process.

% A robust approach for attack variants detection needs to satisfy 2/3 requirements: 1) Effective graph-based matching; 2) Technique-level detection, trade off between TP and FP, In general, the simpler the structure of the feature, the more specific the value is.; 3) Automatic knowledge aggregation;
%  present two real-world attack cases with half of attack techniques replaced. 

}

\vspace{-0.1in}
\section{Conclusion}
\vspace{-0.06in}

We propose a viable solution for retrieving structured threat intelligence from CTI reports in this paper. We use the notion of technique template to identify and aggregate technique-level threat intelligence across massive reports, and leverage the knowledge contained in templates to enhance the attack graph extracted from a CTI report and generate the TKG that introduces the report with enhanced knowledge. We implement our prototype system, AttacKG, and evaluate it with 1,515 real-world CTI reports. Our evaluation results show that \sysname can extract attack graphs from reports accurately and aggregate technique-level threat intelligence from massive CTI reports effectively.

% To facilitate follow-up research, we release the source code of AttacKG at \url{https://github.com/li-zhenyuan/Knowledge-enhanced-Attack-Graph}.

% incorporating data provenance into commercial EDR tools. We use the notion of tactical provenance to reason about causally related threat alerts, and then encode those related alerts into a tactical provenance graph (TPG). We leverage the TPG for risk assessment of the EDR-generated threat alerts and for system log reduction. We incorporated our prototype system, RapSheet, into the Symantec EDR tool. Our evaluation results over an enterprise dataset show that RapSheet improves the threat detection accuracy of the Symantec EDR. Moreover, our log reduction technique dramatically reduces the overhead associated with long-term system log storage while preserving causal links between existing and future alerts.

% We have proposed AttacKG, a system that aggregate technique-level threat intelligence across CTI reports and technique knowledge graphs extracted from CTI reports with the aggregated knowledge. 
% A comprehensive evaluation demonstrates that AttacKG can extract technique knowledge graphs from CTI reports efficiently and accurately.

% Specifically, we first extract attack-related entities and dependencies from CTI reports to generate attack graphs that summarize attack behaviors.
% Then, we adopt automated extracted technique templates to identify attack techniques in attack graphs with a revised graph alignment algorithm.
% Moreover, the information in aligned entities will be extracted to feedback to the templates.

\appendix
\section*{Appendix}

% \section{Entities List}

% \begin{table}[]
%     \centering
%     \vspace{-0.15in}
%     \caption{Attack-relevant Entities (Both IoCs and non-IoCs)}
%     \label{tab:entity_types}
%     \begin{adjustbox}{width=0.76\textwidth}
%     \begin{tabular}{ccl}
%     \toprule
%     Entity Types    & \makecell{Subject\\/Object} & Entity Terms                                        \\ \midrule
%     Actor           & Subject                     & apt, cobalt, rat, actor, lazarus, jrat, \dots       \\
%     Executable      & Subject                     & command, malware, script, payload, vbs, \dots       \\
%     File            & Object                      & files, directory, document, attachments, \dots      \\
%     Network         & Object                      & c2, http, remote, email, server, host, links, \dots \\
%     Registry        & Object                      & registry, keys.                               \\
%     Vulnerability   & Object                      & vulnerability, exploit.                       \\
%     Others          & Object                      & service, credentials, account, wmi, task, \dots     \\ \bottomrule
%     \end{tabular}
%     \end{adjustbox}
%     \vspace{-0.15in}
% \end{table}

\highlightb{

% \vspace{-0.06in}
\section{Another Case Study}
\label{appendix:casestudy}
\vspace{-0.1in}

\textbf{TKGs for attack variants detection.}
As discussed in Section~\ref{subsec:mutation}, frequent and widely used attack variants are posing challenges for detection.
Take a simple \textit{T1204-User Execution} and \textit{T1547-Boot Autostart} two-stage attack excerpted from Frankenstein Campaign, for example. Subfigures (A) and (B) in Figure~\ref{fig:NewCaseStudy} demonstrate the attack and its variants with three nodes mutated. 
Specifically, the file server URL in \textit{T1204} was changed, and the implementation of \textit{T1547} was switched from (A) \textit{Registry Run Keys} to (D) \textit{DLL Side-loading}.
It is noteworthy that such changes will not affect the functionality of the attack.

\begin{figure*}[ht]
    \centering
    \vspace{-0.2in}
    \includegraphics[width=0.98\textwidth]{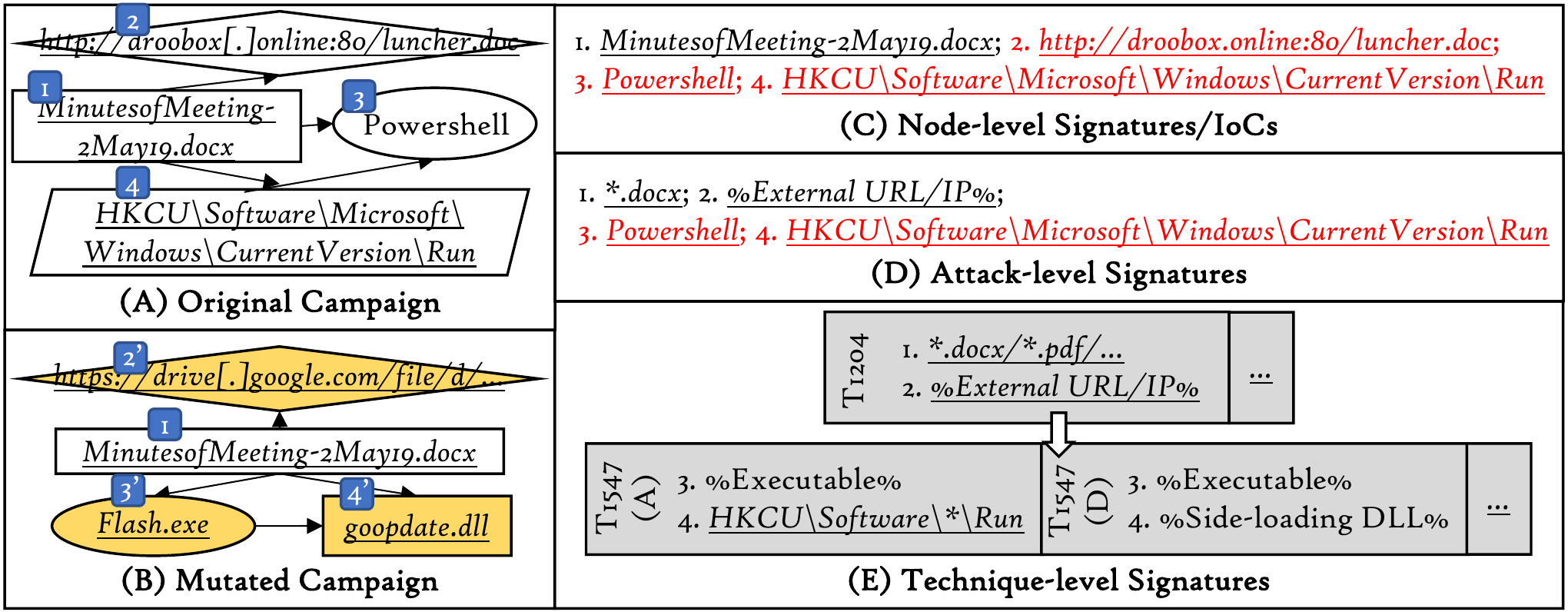}
    \vspace{-0.1in}

    \caption{\highlightb{TKG for attack avariants detection. (\textcolor{red}{Red} outlined mismatch items.)}}
    \label{fig:NewCaseStudy}
    \vspace{-0.25in}
\end{figure*}

Then, three representative intelligence-based detection schemes of different granularity are selected for comparison, namely, 
1) iACE~\cite{Liao} that automatically extracts node-level intelligence (subfigure (C)), 
2) Poirot~\cite{Milajerdi2019} that adopt manually extracted attack graph for threat detection (subfigure (D))
and 3) our approach that aggregate technique-level intelligence from multiple reports (subfigure (E)).
As the matching results shown, to avoid introducing excessive false positives, node intelligence-based detection requires exact matching, which can be easily bypassed by obfuscation and other methods.
By considering structure information, attack-level matching allows the generalization of node information to improve detection generality.
However, attackers can still easily bypass such detection by changing any techniques used in the campaign.

Nevertheless, the technique-level intelligence we provide enables detectors to detect different attack techniques independently. 
Moreover, the pooled technique knowledge from multiple reports can effectively improve the detection of various variants. 
And the aggregated intelligence can be automatically merged with approaches like Eiger~\cite{Kurogome2019} for better generality.

}

\highlightb{
\vspace{-0.1in}
\section{Selecting the Threshold Value} 
\label{sec:threshold}
\vspace{-0.1in}

The selection of the threshold value for node/graph alignment scores affects the accuracy and efficiency of AttacKG. Specifically, too low a threshold for graph alignment score could result in premature matching (false positives), while too high could lead to missing reasonable matches (false negatives). For node alignment score, too low a threshold could leave unnecessary alignment candidates and cost longer report analysis time, while too high could lead to false negatives. Thus, there are trade-offs in choosing optimal threshold values. To determine optimal threshold values, we measure the F-score and report analysis time using varying threshold values, as shown in Figure \ref{fig:threshold}, and select optimal threshold values (0.65 for node alignment, 0.85 for graph alignment) that make each index better at the same time. 
% This analysis is done based on the highest alignment score calculated in 20 iterations of Poirot’s search algorithm for all the attack and benign scenarios we have evaluated. As it is shown, the highest F-score value is achieved when the threshold is at the interval [0.17, 0.54], which is the range in which all attack subgraphs are correctly found, and no alarm is raised for benign datasets. The middle of this interval, i.e., 0.35, maximizes the margin between attack and benign scores, and choosing this value as the optimal threshold minimizes the classification errors,. Therefore, we set the Cthr to 3 which results in 1 Cthr = 1 3 which is close to the optimal value.

\begin{figure*}[ht]
\vspace{-0.2in}
     \centering
     \begin{subfigure}[b]{0.45\textwidth}
         \centering
         \includegraphics[width=\textwidth]{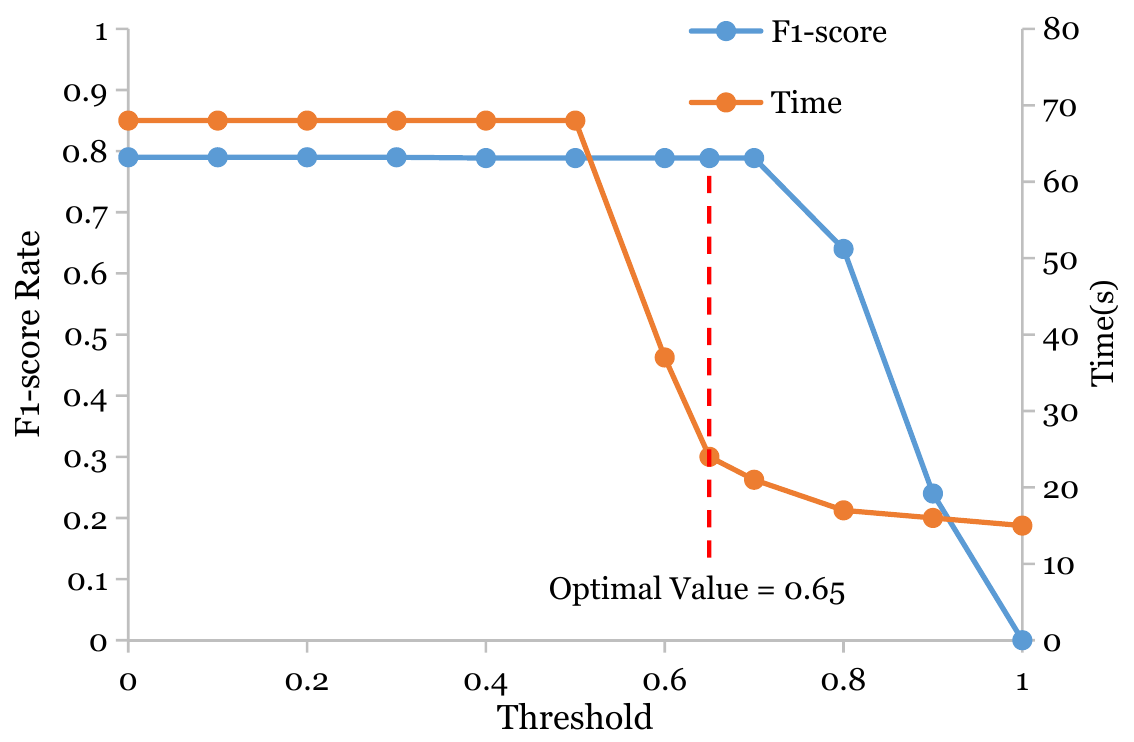}
        \vspace{-0.15in}
         \caption{\small{node alignment score}}
         \label{fig:node_threshold}
     \end{subfigure}
     \hfill
     \begin{subfigure}[b]{0.45\textwidth}
         \centering
         \includegraphics[width=\textwidth]{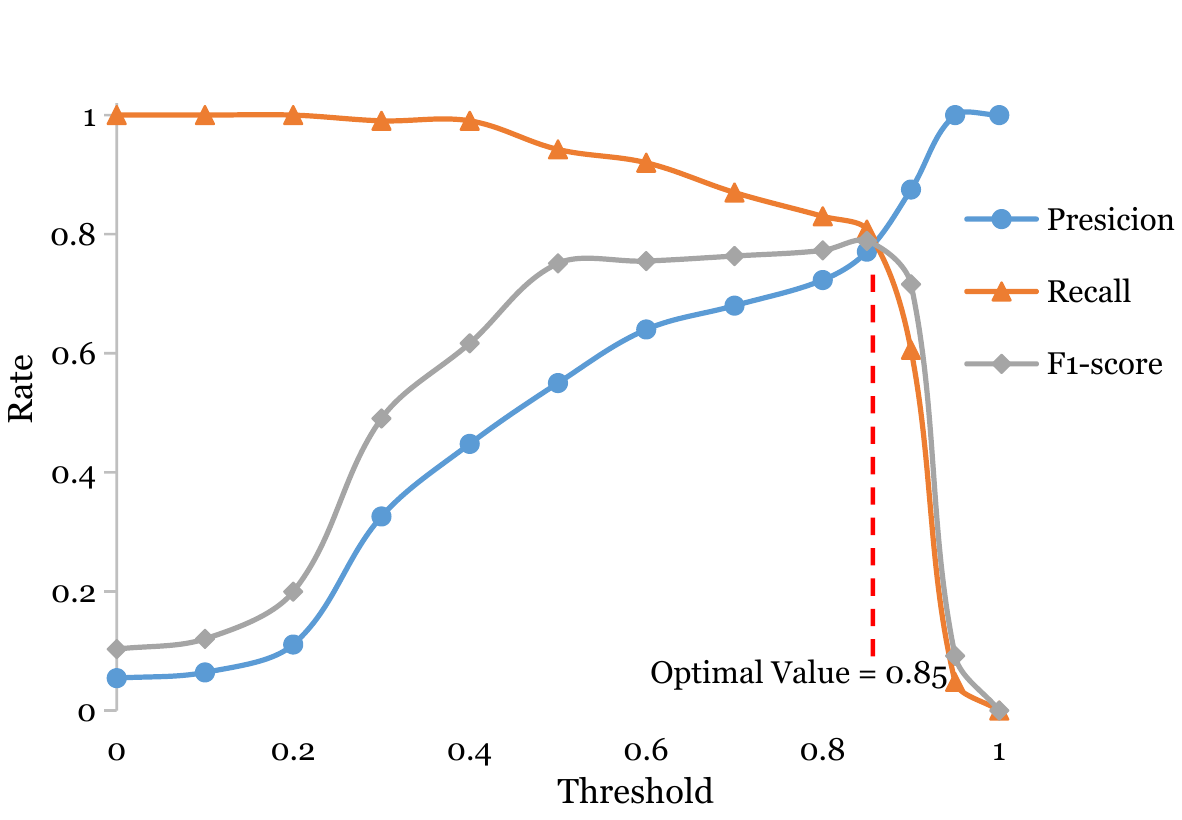}
        \vspace{-0.15in}
         \caption{\small{graph alignment score}}
         \label{fig:graph_threshold}
     \end{subfigure}
     \vspace{-0.1in}
    \caption{\highlightb{\small{Threshold selection}}}
    \label{fig:threshold}
\vspace{-0.2in}
\end{figure*}

}
\vspace{-0.1in}
\section{Ablation Study of AttacKG.} 
\label{sec:ablation_study}
\vspace{-0.1in}

In particular, we first remove part of the attributes in entities: the IoC information and natural language text termed $\sysname_{w \setminus o\ IoC\ information}$ and  $\sysname_{w \setminus o\ natural\ language\ text}$, respectively.
Note that unlike the EXTRACTOR's practice of merging entities, which may result in information loss, we only remove partial entity attributes without sacrificing the structural information of attack graphs.
Moreover, we obtain another variant by filtering out dependencies in attack graphs termed $\sysname_{w \setminus o\ dependencies}$.
That is, we predict attack techniques only based on entity sets.
Finally, we disable the graph simplification component termed $\sysname_{w \setminus o\ graph\ simplification}$.

As different component combinations may affect the distribution of alignment scores, we adjust and choose identification thresholds separately for \sysname variants in light of the optimal F1-scores.
The experimental results are summarized in Table~\ref{tab:breakdown_analysis}.
We find that removing any component would degrade \sysname's overall performance, which well justifies our design choice.
Especially, $\sysname_{w \setminus o\ dependencies}$ consistently performs the worst across all evaluation metrics. 
It verifies the substantial influence of graph structures in technique identification.

% https://www.tablesgenerator.com/latex_tables
\begin{table}[htbp]
\footnotesize
\centering
\vspace{-0.25in}
\caption{Ablation study of different components used in technique identification.}
\label{tab:breakdown_analysis}
% \vspace{-0.1in}
\begin{tabular}{lccc}
\toprule
% \multicolumn{1}{c}{\multirow{2}{*}{}} & \multicolumn{3}{c}{Techniques} \\ \cline{2-4} 
Components        & Precision  & Recall  & F1-Score \\  \midrule
w/ all   component                      & 0.782      & \textbf{0.860}   & \textbf{0.819}    \\ 
w/o IoC information                     & \textbf{0.833}     & 0.600   & 0.698    \\ 
w/o natural language text                 & 0.690      & 0.800   & 0.741    \\ 
w/o dependencies                        & 0.667      & 0.480   & 0.558    \\ 
w/o graph simplification                      & 0.696      & 0.780   & 0.736    \\ \bottomrule
\end{tabular}
\vspace{-0.25in}
\end{table}

\highlightb{

\vspace{-0.1in}
\section{Efficiency of AttacKG}
\vspace{-0.1in}

\textbf{Setup.} We experimentally compared AttacKG's efficiency with TTPDrill and Extractor on the 16 CTI report samples mentioned in Section~\ref{subsec:eval_setup} on a PC with AMD Ryzen 7-4800H Processor 2.9 GHz, 8 Cores, and 16 Gigabytes of memory, running Windows 11 64-bit Professional. % with average 278.15 words length. The experiment results are obtained using 
The size of the reports used as samples ranges from 61 words to 1029 words, with an average of 278.2 words. 

\textbf{Results.} Among the three models, Extractor adopts the most complex system consisting of multiple NLP models and has the highest runtime overhead, taking 239.70 seconds on average to parse a report. 
While AttacKG simplified the graph extraction process, making it possible to identify attack techniques and the extraction of attack graph and still utilize much less time.
On average, graph extraction takes 8.9 seconds, and technique identification takes 15.1 seconds, totaling 24.0 seconds. 
TTPDrill, on the other hand, uses the simplest model without constructing attack graphs and is therefore also the fastest, taking only 5.9 seconds on average to analyze a report, but at the cost of a high false-positive rate.
}

\bibliographystyle{splncs04}
\bibliography{AttacKG}

\end{document}